\documentclass[usenatbib]{mnras}

\usepackage[T1]{fontenc}
\usepackage{ae,aecompl}

\usepackage{verbatim} 

\usepackage{graphicx}	
\usepackage{amsmath}	
\usepackage{amssymb}	

\title[Detection of IMBHs in GCs using integrated-light spectroscopy]{Prospects for detection of intermediate-mass black holes in globular clusters using integrated-light spectroscopy}
\author[R. de Vita et al.]{
R. de Vita,$^{1}$\thanks{E-mail: r.devita@student.unimelb.edu.au}
M. Trenti,$^{1}$
P. Bianchini,$^{2}$
A. Askar,$^{3}$
M. Giersz,$^{3}$
G. van de Ven$^{2}$
\\
$^{1}$The University of Melbourne, School of Physics, VIC 3010, Australia\\
$^{2}$Max-Planck Institute for Astronomy, Koenigstuhl 17, 69117 Heidelberg, Germany\\
$^{3}$Nicolaus Copernicus Astronomical Centre, Polish Academy of Sciences, ul. Bartycka 18, 00-716 Warsaw, Poland
}

\date{Accepted 2017 February 3. Received 2017 January 30; in original form 2016 October 14}

\pubyear{2016}

\begin{document}
\label{firstpage}
\pagerange{\pageref{firstpage}--\pageref{lastpage}}
\maketitle

\begin{abstract} The detection of intermediate mass black holes (IMBHs) in Galactic globular clusters (GCs) has so far been controversial. In order to characterize the effectiveness of integrated-light spectroscopy through integral field units, we analyze realistic mock data generated from state-of-the-art Monte Carlo simulations of GCs with a central IMBH, considering different setups and conditions varying IMBH mass, cluster distance, and accuracy in determination of the center. The mock observations are modeled with isotropic Jeans models to assess the success rate in identifying the IMBH presence, which we find to be primarily dependent on IMBH mass. However, even for a IMBH of considerable mass (3\% of the total GC mass), the analysis does not yield conclusive results in 1 out of 5 cases, because of shot noise due to bright stars close to the IMBH line-of-sight. This stochastic variability in the modeling outcome grows with decreasing BH mass, with approximately 3 failures out of 4 for IMBHs with 0.1\% of total GC mass. Finally, we find that our analysis is generally unable to exclude at 68\% confidence an IMBH with mass of $10^3~M_\odot$ in snapshots without a central BH. Interestingly, our results are not sensitive to GC distance within 5-20 kpc, nor to  mis-identification of the GC center by less than $2\arcsec$ ($<20\%$ of the core radius). These findings highlight the value of ground-based integral field spectroscopy for large GC surveys, where systematic failures can be accounted for, but stress the importance of discrete kinematic measurements that are less affected by stochasticity induced by bright stars.

\end{abstract}



\section{Introduction}
The existence of intermediate mass black holes (IMBHs) with masses between few $M_\odot$ (stellar black holes of $\approx 100\ M_\odot$) and billions of $M_\odot$ (supermassive black holes of $\approx 10^6\ M_\odot$) is of particular interest, especially in the context of the formation and evolution of galaxies and dense stellar systems such as globular clusters (GCs). The natural extension of the well-known $M-\sigma$ relation for galaxies suggests that the typical central velocity dispersions in GCs might be associated to the presence of IMBHs with masses of $10^{3-4}\ M_\odot$ (see, e.g., \citealt{ferrarese:00}, \citealt{gebhardt:00}). To support this extrapolation, several scenarios for the formation of such objects have been proposed, including run-away collapse of massive stars \citep{portegies-zwart:04}, early-time accretion of ejecta from asymptotic giant branch stars in the context of multiple stellar population formation \citep{vesperini:10}, dynamical interactions of hard binaries \citep{giersz:15}, or possibly seeding from massive Population III stars if the oldest globular clusters form during the epoch of reionization at redshift $z\sim 8-10$ \citep{trenti:15,ricotti:16}. 

$\ $ 

In recent years there have been many studies focused on searching for these objects. Some tentative detections have been claimed, but a non-controversial proof of their presence is still lacking (e.g., \citealt{haggard:13} and references therein). Some of the observational techniques used are based on the detection of the radio and X-ray emission associated with accretion processes, but these are complicated by the lack of gas in the old globular clusters in our Galaxy (see, e.g, \citealt{farrell:12}, \citealt{mezcua:13} ). Stellar dynamics, and in particular modeling of the central velocity dispersion profile is another tool for searching IMBHs, but measurements are very challenging because the sphere of influence of the BH is limited to a few arcsec, even for the closest and most massive GCs such as $\omega$ Cen (see  \citealt{noyola:10}; \citealt{vandermarel:10}). 
Finally, the fact that these events are expected to be also sources of gravitational radiation promotes the interferometers such as advanced-LIGO as further instruments to search for IMBHs (see, e.g., \citealt{mandel:08}, \citealt{konstantinidis:13}, \citealt{macleod:16}).

A complementary tool to approach the problem is that of identifying novel dynamical signatures for the presence of IMBH in globular clusters based on numerical modeling of globular cluster dynamics in presence of an IMBH. Starting from initial direct N-body simulations more than a decade ago \citep{baumgardt:04a,baumgardt:04b,trenti:07}, simulations have progressed significantly, and are now approaching realistic particle numbers with direct integration algorithms that include post-newtonian corrections (e.g. \citealt{macleod:16,wang:16}), and routinely include more than one million particles through Monte Carlo methods (\citealt{giersz:15}; see also \citealt{rodriguez:15}). These investigations have shown that a central massive black hole is expected to induce the formation of a shallow cusp in the projected surface brightness and to prevent the core collapse by enhancing three-body interactions within its sphere of influence (see, e.g, \citealt{baumgardt:05}). In addition, it has been shown that the IMBH is able to quench the process of mass segregation (see e.g., \citealt{gill:08}, \citealt{pasquato:09}, \citealt{pasquato:16}). However, one important caveat is that these signatures may be only necessary but not sufficient conditions to infer the presence of an IMBH, because other dynamical processes could mimic them (see, e.g., \citealt{hurley:07}; \citealt{trenti:10}; \citealt{vesperini:10b}).

Recently, the majority of the observational claims about the presence of IMBHs comes from kinematic measurements in the inner core of Galactic GCs. Kinematic observations suggesting the presence of IMBHs are traditionally based on the search for a rise of the central velocity dispersion. This method requires both high spatial resolution, to resolve the very crowded central region of GCs (few central arcseconds), and very precise velocity measurements with accuracy $\approx1$ km s$^{-1}$. 

So far, the available observations of the central regions of Galactic GCs have led to contradictory results when applied to the same object in a few instances (e.g., \citealt{noyola:10}, \citealt{vandermarel:10}, \citealt{lutzgendorf:13}, \citealt{lanzoni:13}, \citealt{lutzgendorf:15}). In general, two different strategies are used in order to infer the presence of IMBHs: resolving individual star velocities (line-of-sight velocities or proper motions) or using unresolved kinematic measurements, for example with integral field unit (IFU) spectroscopy. 
Both these methods suffer technical difficulties in obtaining the critically needed kinematic measurements in the very center of the system (e.g., the problem of shot-noise for integrated-light measurements and the effects of crowding for line-of-sight velocities and proper motions).  
In particular, integrated-light spectroscopy tends to detect rising central velocity dispersions, suggesting the presence of IMBHs (see for example, \citealt{noyola:10} for $\omega$ Cen, or \citealt{lutzgendorf:11} for NGC 6388), while resolved stellar kinematics are consistent with a flat velocity dispersion profile, that is no massive black hole (see \citealt{vandermarel:10} for proper motion measurements of $\omega$ Cen, and \citealt{lanzoni:13} for discrete line-of-sight measurements in NGC 6388). However, in a few other cases where both discrete and integrated-light profiles are available for the inner $10\arcsec$, the observational methods agree (e.g., see NGC 2808, NGC 6266, NGC 1851 in \citealt{lutzgendorf:13}).

For both unresolved and resolved kinematics, the constraints on the IMBH mass are generally determined by fitting the observed velocity dispersion profiles with different families of Jeans models (e.g, \citealt{vandermarel:10}). These models are typically constructed by making assumptions on the mass-to-light ratio profile $M/L(r)$ in order to calculate the intrinsic mass distribution of the luminous component from the surface brightness profile. The velocity dispersion profile is then calculated by solving the Jeans equation for hydrostatic equilibrium in a spherical stellar system (see e.g., \citealt{bertin:14}). Besides the Jeans modeling, other analysis techniques used include the Schwarzschild's orbit superposition method used in \citealt{vandeven:06} or a method in which the fit of the observed velocity dispersion profiles is performed using a grid of N-body simulations (see e.g., \citealt{jalali:12}; \citealt{baumgardt:17}).

The main goal of this work is to characterize under which conditions (IMBH mass, GC distance, accuracy in the determination of the center) the integrated-light IFU data are able to measure accurately the mass of the IMBH, as inferred from realistic mock observations of simulated star clusters with a central IMBH. By means of the software SISCO developed by \citealt{bianchini:15}, we are able to create mock IFU observations of the central regions of GCs. The set of observations is produced starting from a set of Monte Carlo cluster simulations (MOCCA simulations by \citealt{giersz:15,askar:16}; see also \citealt{askar:17}, for a similar application of SISCO to MOCCA simulations) that include a range of different IMBH masses (from $0$ to $10^4\ M_\odot$). In order to quantify the significance of a central rise in the simulated velocity dispersion profiles, we fit these profiles with a one-parameter family of isotropic Jeans models. In this way, we are able to estimate quantitatively and objectively the IMBH mass and, thus, to directly test the ability of the observations to successfully recover the mass of the central black hole.

The paper is organised as follows. In Sect.\ref{sect.2} we present the set of simulations used and we briefly describe the SISCO code used to produce  the mock IFU observations. Moreover, we describe the dynamical models used to fit the observed profiles. In Sect.\ref{sect.3} we present the results of our analysis and in Sect.\ref{sect.4} we give our conclusions.

 \section{Methods}
 \label{sect.2}
 \subsection{MOCCA simulations}
 \label{simulations}

In this work we resort to Monte Carlo simulations of GCs that include the presence of a central IMBH.
These simulations are part of about 2000 GC models run in the framework of the MOCCA SURVEY I project (see \citealt{askar:16} for a description of the Survey).
The IMBH in the simulated clusters is formed dynamically from stellar-mass BH seeds as a result of dynamical interactions and mergers in binaries. 

All models have a stellar initial mass function (IMF) given by \citet{kroupa:01} with minimum and maximum stellar masses taken to be $0.08\ M_\odot$ and $100\ M_\odot$, respectively. 
Supernovae (SN) natal kick velocities for neutron stars and BHs were drawn from a Maxwellian distribution with a dispersion of 265 km/s \citep{hobbs:05}. 
For most models, natal kicks for BHs were modified according to the mass fallback procedure described by \citet{belczynski:02}. To model the Galactic potential, a point mass approximation with the Galaxy mass equal to the mass enclosed inside the cluster Galactocentric distance is assumed. Additionally, it is also assumed that all clusters have the same rotation velocity, equal to 220 km/s. So, depending on the cluster mass and tidal radius the Galactocentric distances span from about 1 kpc to about 50 kpc. 

Here, we selected a subsample of MOCCA runs for analysis, and report in Table \ref{tab:SIM} the key properties at time $t=12$ Gyr. In addition to this snapshot, we also consider three additional snapshots at 11.7, 11.8, 11.9 Gyr to assess the robustness of our conclusions against variance introduced by a different dynamical state of a system of otherwise similar global properties. 

 \begin{table*}
	\centering
	\caption{Set of MOCCA simulations, labeled S0-S5, used in this paper and taken from MOCCA-SURVEY Database I  \citet{askar:16}. For each simulation we report the quantities relative to the snapshot at 12 Gyr: number of stars $N$; total mass $M$ and IMBH mass $m_\bullet$ (solar units); binary fraction $f_b$; projected truncation radius $R_t$, projected core radius $R_c$ (from the surface brightness profile), projected half-light radius $R_h$ and intrinsic radius for the IMBH sphere of influence $r_{\bullet}$ (pc); concentration parameter $C=\log(R_t/R_c)$.}
	\label{tab:SIM}
	\begin{tabular}{lcccccc}
		\hline
					&S0				& S1				& S2				&S3				& S4				&S5\\	
		\hline
		$N$			&$1.1\times 10^6$	&$1.0\times 10^6$	&$1.0\times 10^6$	&$2.9\times 10^5$	&$5.6\times 10^5$	&$6.3\times 10^5$\\
		$M$ 			&$3.0\times 10^5$	&$3.0\times 10^5$	&$3.4\times 10^5$	&$9.6\times 10^4$	&$1.8\times 10^5$	&$2.0\times 10^5$\\
		$m_{\bullet}$	&$1.0\times 10^4$	&$9.9\times 10^3$	&$5.6\times 10^3$	&$1.6\times 10^3$	&$2.3\times 10^2$	&- \\
		$f_b$		&7\%				&8\%				&4\%				&5\%				&5\%				&3.6\%\\	
		$R_t$		&102.6			&80.9			&89.6			&59.8			&69.1			&89.1\\
		$R_c$		&0.3				&0.4				&0.2				&0.4				&0.6				&1.3\\
		$R_h$		&1.9				&2.1				&2.6				&2.2				&2.0				&2.3\\
		$r_{\bullet}$	&0.452			&0.559			&0.198			&0.260			&0.012			&-\\	
		$C$			&2.53			&2.30			&2.64			&2.17			&2.06			&1.8\\
		\hline
	\end{tabular}
\end{table*}

\subsection{SISCO software}
The software SISCO (Simulating Stellar Cluster Observation) produces a mock IFU data cube starting from a simulated star cluster (for a detailed description see \citealt{bianchini:15}). The software derives a medium-high resolution spectrum ($R\approx20000$) in the wavelength range of the Ca triplet (8400-8800 \r{A}) for every star of the simulation, based on mass, effective temperature, luminosity and metallicity. The spectral coverage and resolution are tailored to mock observations of typical IFU instruments like FLAMES@VLT in ARGUS mode \citep{pasquini:12}. SISCO allows users to define specific instrumental setups: we fix the size of the field-of-view (FOV) to $20\times20$ arcsec$^2$ and the spaxel scale to $0.25$ arcsec; we adopt a Moffat shape for the point spread function with seeing condition of $1$ arcsec and shape parameter $\beta=2.5$. Finally, we mimic an observation with an average signal-to-noise ratio of $S/N\simeq10$ per \r{A} (for a discussion on the fixed values of the parameters used in our mock observations, see \citealt{bianchini:15}). In order to simulate different observing conditions, we change three parameters: the distance to the cluster, the direction of its projection in the sky, and optionally introduce an off-set between the centre of the simulated IFU field and the centre of the cluster, to reflect the uncertainty in determining the centre of an observed GC. The final output of the code is a three-dimensional data cube in which each spatial pixel has an assigned spectrum.

\begin{figure*}
\resizebox{\hsize}{!}{\includegraphics[clip=true]{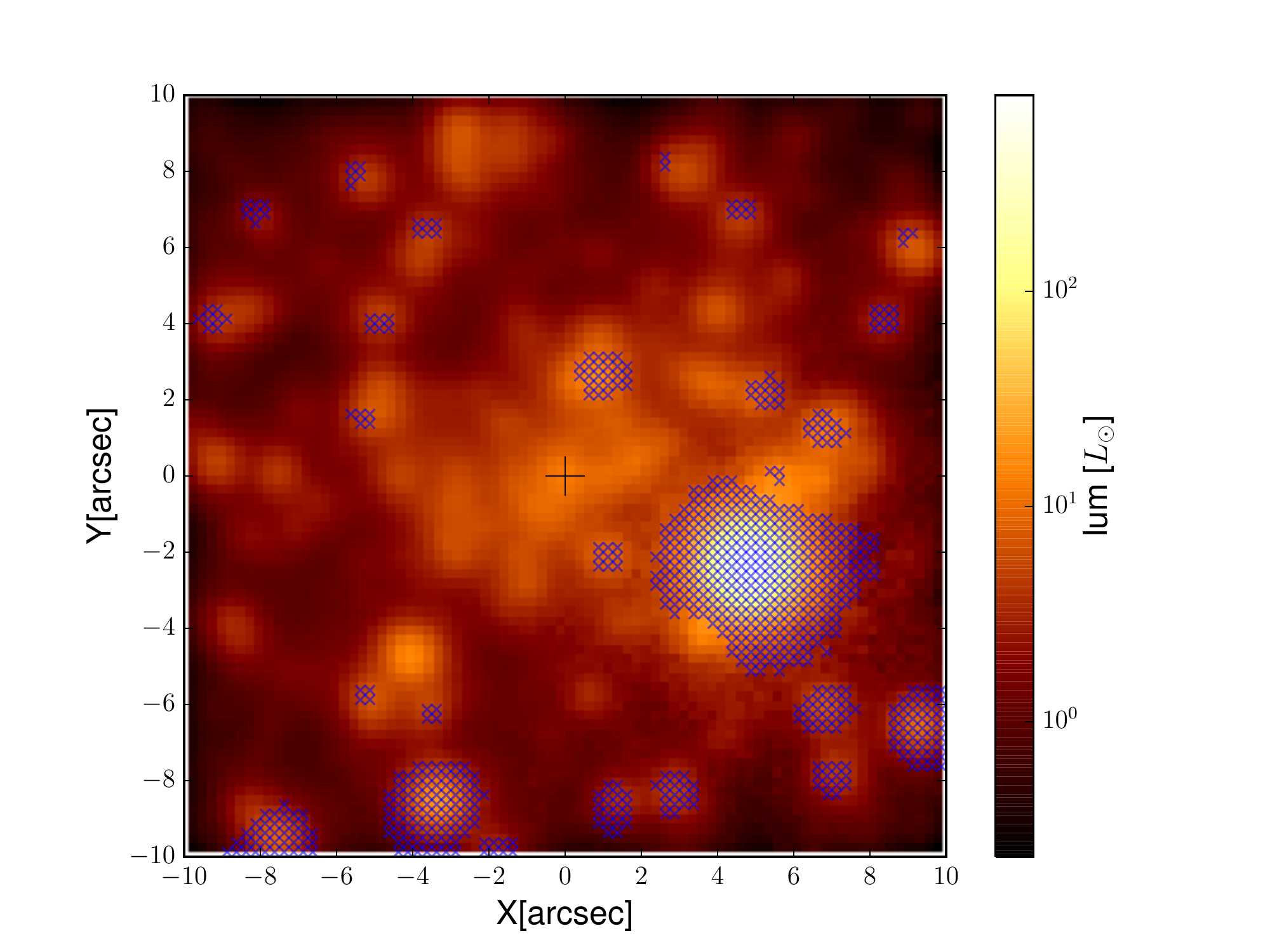}
\includegraphics[clip=true]{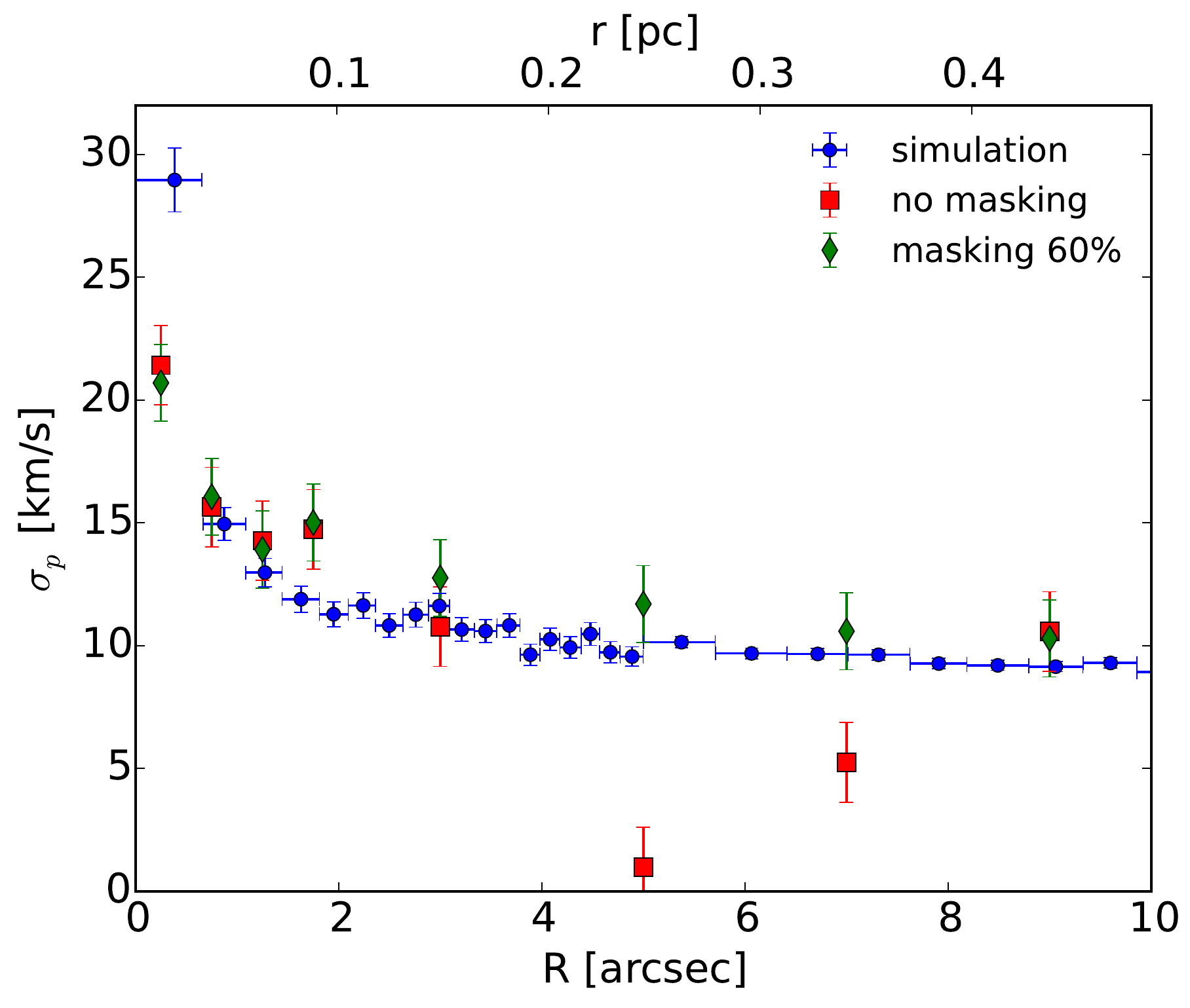}
}
\caption{
{\it Left panel:} luminosity map of the simulated cluster S0 observed at a distance of 10 kpc with a FOV around the centre of the cluster. The masked spaxels are indicated with the `x' symbols. 
{\it Right panel:} Central velocity dispersion profiles for the same cluster. The blue circles show the profile measured directly from the MOCCA simulation considering objects with masses between $0.7\ M_\odot$ and $1\ M_\odot$ and using the barycentre velocity for binary systems. The profile inferred from analysis of the mock IFU data cube is plotted in red squares without masking, and in green diamonds with the bright-object masking procedure discussed in the text.
}
\label{LM+VDP}
\end{figure*}

\subsection{Determination of the velocity dispersion profile}
\label{subsect:vel disp prof}
In order to mimic real observation as closely as possible, we construct the observed velocity dispersion profile by integrating our mock IFU data available for the region inside the FOV, and combine it with a line-of-sight velocity dispersion obtained for the outer parts of the system. 
The outer profile is obtained directly from the simulation by using only the velocities of the red giant stars, which are those generally used for resolved kinematics from the ground. For this analysis, we treat the binary stars as single objects with the velocity of their center of mass. 
For the inner profile, once the IFU data cube is simulated through SISCO, we divide the FOV in radial bins, summing the spectra in each bin, with the aim of interpreting the data cube through a spherical dynamical model. The binned spectra are analysed with the pPXF code (\citealt{cappellari:04}) to derive the velocity dispersion (and the corresponding error) from line broadening. 

As highlighted in \cite{bianchini:15}, when integrated-light measurements are used, the presence of a few bright stars can introduce systematic effects in the reconstruction of the observed velocity dispersion profile.  For this reason, we introduced masking of the brightest sources. Specifically, we exclude from the analysis the spaxels in which the contribution of a single stars exceeds the 60\% of the total luminosity (we adopt the same percentage used in \citealt{lutzgendorf:13}). This information is provided directly by the simulation, thus, from an observational point of view, we are considering an ideal case scenario. In Fig.~\ref{LM+VDP} we show the luminosity map (left panel) and the radial velocity dispersion profile (right panel) for the central region of simulation S0 at a distance of $10$ kpc. The velocity dispersion profile constructed from the simulation (blue circles) is obtained by considering objects with mass in the range $0.7-1\ M_\odot$ to mimic the average (luminosity weighted) mass in the FOV and by using the barycentre line of sight velocities for the binary stars to avoid great scatter in the profile (the binning must be fine in order to sample the central region). The velocity dispersion profile obtained by masking the IFU data over the regions shown as green diamonds is consistent with that constructed directly from the simulation. Without the masking, there is an evident discrepancy in the range $4-8$ arcsec between the observed profile and the expected one. The presence of a very bright star in this radial bin thus influences the observed velocity dispersion with the tendency of underestimating the measurement. 

Despite the masking of the brightest sources, the observed profiles tends to slightly overestimate the velocity dispersion for radii $>2\arcsec$. 
This effect is possibly due to the presence of hard binaries which may influence the velocity dispersion determination. Indeed, the observed velocity associated to a binary system could largely exceed the mean field velocity because of the high-speed orbital motions. This effect is merely an observational feature associated to line-of-sight velocity dispersion measurements, and in principle it could be accounted for if proper-motion kinematic is available (see e.g., \citealt{bianchini:16b}), or through theoretical modeling of the binary population, both in energy and position space. However, this investigation is beyond the scope of the present paper and we limit our analysis to include the effects of the population of binaries into the construction of our mock observations.

\begin{figure}
\resizebox{\hsize}{!}{\includegraphics[clip=true]{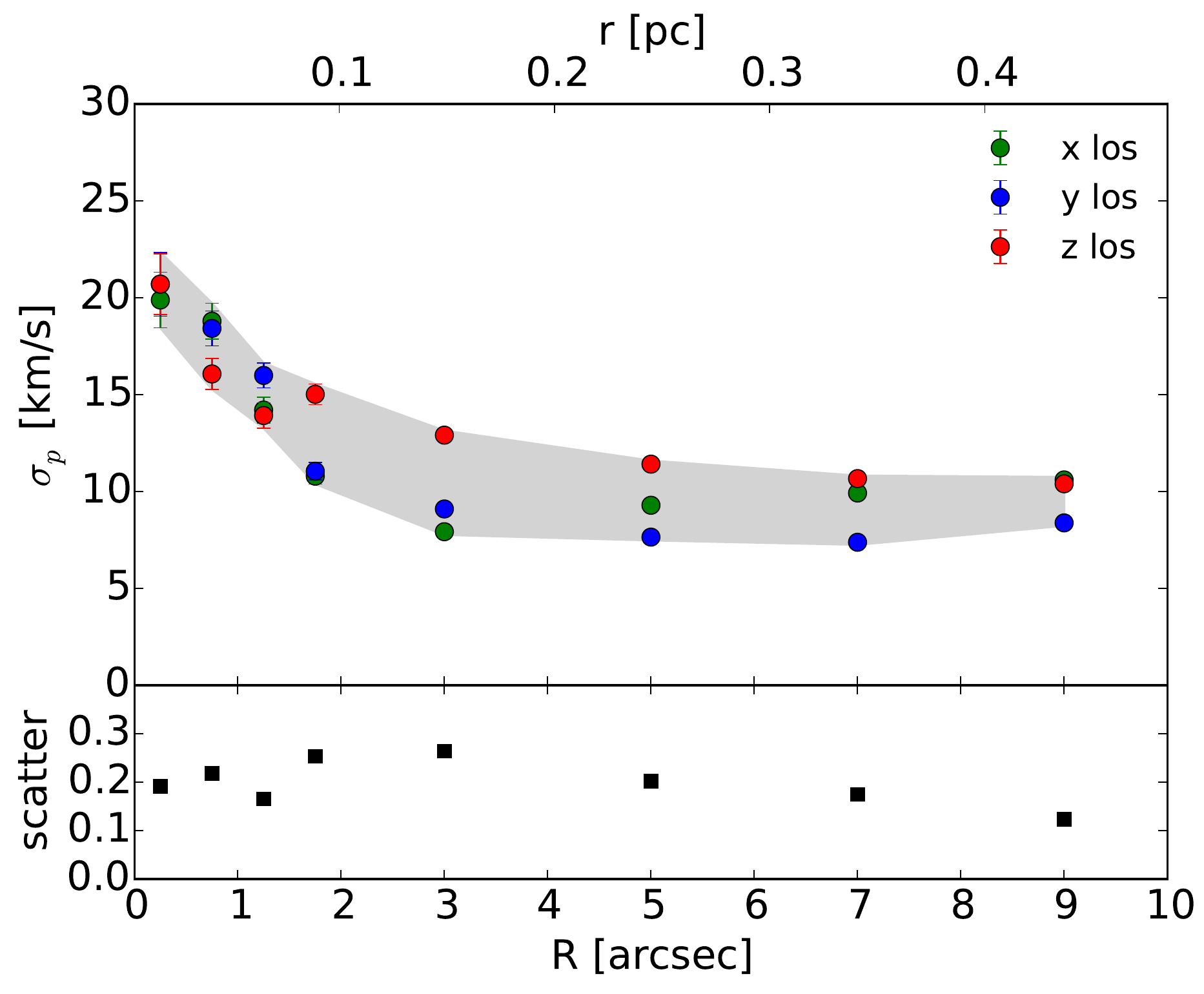}
}
\caption{Observed velocity dispersion profiles of the cluster S0 at a distance of 10kpc for three different directions of the line-of-sight. The bottom panel shows the scatter in the velocity dispersion, that is, the difference of the greater and the lower value for each radial position divided by an average central value for the velocity dispersion. 
}
\label{los dir}
\end{figure}

Finally, we produced different realisations of the same simulation to test the intrinsic scatter of the velocity dispersion profile. In particular, we changed the direction of the line-of-sight for the mock observation of the cluster S0 under canonical conditions (that is, at 10 kpc and with the FOV pointing to the centre). For three different projections of the simulated cluster we obtained a velocity dispersion profile for which the scatter is uniform along the entire profile and it does not exceed the 30\% of the central value (see Fig.~\ref{los dir}). Especially for the outer points, where the signal is stronger, the intrinsic scatter is much larger than the errors calculated by the pPXF software from line broadening. For this reason, for the rest of our analysis, we will consider an error $\delta \sigma$ for all the points in the observed profile calculated by considering the error $\delta \sigma_0$ obtained  by pPXF for the innermost point. In particular, the error for any outer point is given by $\delta \sigma=\delta \sigma_0(\sigma/\sigma_0)$, where $\sigma_0$ and $\sigma$ are the velocity dispersions of the innermost point and the outer point, respectively.

\subsection{Dynamical modeling}
As usually done in the literature, we fit the velocity dispersion profile derived from the mock observations with a family of dynamical models in which the IMBH mass is treated as a free parameter. 
We place ourself under the ideal conditions of assuming a perfect knowledge of the spherically symmetric distribution of the stellar particles. Thus, we adopt a spherical and isotropic Jeans model in which the total gravitational potential of the system, $\Phi(r)$, is given by the sum of the stellar/remnant contribution $\Phi_*(r)$ and the IMBH contribution $\Phi_{\bullet}(r)=-m_{\bullet}/r$ (we fixed the gravitational constant G=1). 
The mass distribution of stars and remnants is directly inferred from the simulation. \footnote{Note that this approach is different from what usually done in real observations, for which the inferred mass distribution can be affected by observational biases or specific assumptions on the mass-to-light ratio profile.} The density $\rho$ at each radius $r$ is estimated using spherical cells, by dividing the total mass (including the IMBH in the innermost cell) of the particles by the shell volume. The radial profile $(d\Phi_*/dr)(r)$ follows from the Gauss theorem as
\begin {equation}
\frac{d\Phi_*}{dr}(r) = \frac{M(<r)}{r^2},
\end{equation}
where $M(<r)$ is the stellar and remnants mass (excluding the central IMBH) contained in the sphere of radius $r$.

We then calculate the intrinsic velocity dispersion profile $\sigma(r)$ from the spherical isotropic Jeans equation
\begin{equation}
\frac{1}{\rho}\frac{d}{dr}\left(\rho\sigma^2\right)=-\frac{d\Phi}{dr},
\end{equation}
which has the solution
\begin{equation}
\label{VDP_intr}
\sigma^2(r)=\frac{1}{\rho(r)}\int_r^\infty \rho(r')\frac{d\Phi}{dr}(r')\ dr'.
\end{equation}
By defining the first derivative of the gravitational potential as
\begin{equation}
 \frac{d\Phi}{dr}(r)=\frac{m_{\bullet}}{r^2}+\eta^2 \frac{d\Phi_*}{dr}(r),
 \end{equation}
where $\eta$ is a constant, the total velocity dispersion in Eq.~\eqref{VDP_intr} can be explicitly written as the sum of the IMBH contribution and the stellar/remnant contribution.
In this way, for a given mass density profile, the velocity dispersion profile depends only on the mass $m_{\bullet}$ of the central IMBH and on $\eta$, which are free parameters of the model. Note that the second parameter $\eta\sim 1$ (a global multiplicative rescaling of the velocity dispersion) has been added in order to take into account two aspects of the kinematics. The first and most relevant is related to possible observational biases due to the presence of binary stars. The effect of considering individual motions in binary systems should be that of increasing the observed velocity dispersion, as orbital speeds may arbitrarily differ from the average field velocity. The second aspect is related to the presence of (partial) energy equipartition. In fact, the velocity dispersion of a population of kinematic tracers will in general depend on their characteristic mass compared to the average particle mass in the system, as well as from the dynamical state of the system, with realistic mass spectra achieving only a partial equipartition \citep{baumgardt:03,trenti:13,bianchini:16}. For these considerations we find convenient to limit $\eta$ to the plausible range $[0.5,2.0]$.

Finally, for a proper comparison with the observation we consider the projected velocity dispersion profile $\sigma_p(R)$ obtained by integrating the density-weighted intrinsic profile along the line-of-sight direction: 
\begin{equation}
\label{VDP_proj}
\sigma_p(R)=\left[\frac{\int_R^{\infty} r(r^2-R^2)^{-1/2} \rho(r)\sigma^2(r)\ dr}{\int_R^{\infty}  r(r^2-R^2)^{-1/2}  \rho(r)\ dr}\right]^{1/2}.
\end{equation} 
 
\begin{figure}
\resizebox{\hsize}{!}{\includegraphics[clip=true]{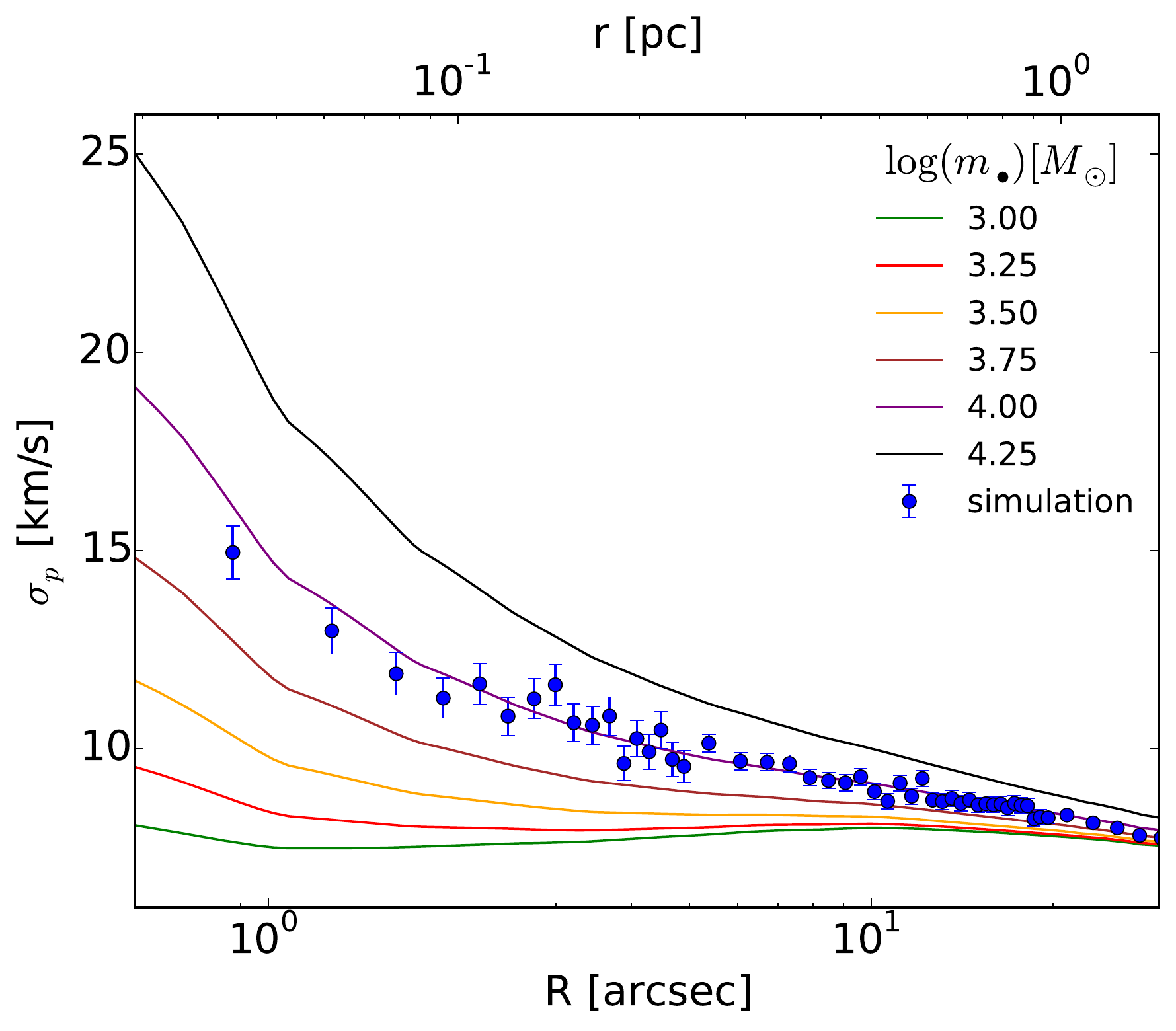}
}
\caption{
Projected velocity dispersion profiles in the centre of the cluster of the cluster in simulation S0 considered at a distance of $10$ kpc. The solid lines represent six velocity dispersion profiles derived from different Jeans model obtained by varying the IMBH mass from $10^3$ to $10^{4.25}\ M_\odot$ with a logarithmic separation of 0.25. By increasing the mass, the central peak becomes steeper. The circles represent the velocity dispersion profile derived directly from the simulation considering objects with masses between $0.7$ and $1M_\odot$. 
}
\label{JMs}
\end{figure}

In Fig.~\ref{JMs} we show the dynamical modeling predictions for the velocity dispersion profiles obtained by changing the mass of the central IMBH for the simulation S0 (we fixed $\eta=1$). As reference, we plot also the velocity dispersion profile constructed directly from the simulation for object in the mass range $[0.7-1.0M_\odot]$, that is the range that includes the average mass observed in our FOV. It is noteworthy that there is good agreement between the inner profile and the Jeans model profile with an IMBH of $10^4\ M_\odot$, which is the actual value for this simulation. 

To quantify the recovery of the IMBH mass we carry out a maximum likelihood fit by minimizing the two-dimensional chi-square function
\begin{equation}
\label{chi2}
\chi^2(m_\bullet,\eta)=\sum_{i=1}^{N}\left[ \frac{\sigma_{obs}(R_i)- \sigma_p(R_i;m_\bullet,\eta)}{\delta \sigma(R_i)} \right]^2,
\end{equation}
where $\sigma_{obs}(R_i)$ are the observed velocity dispersion values (with the error $\delta \sigma(R_i)$) for the $N$ radial bins in which the FOV has been divided (see the previous subsection for a proper description of the error $\delta \sigma(R_i)$).

%
\section{Results}
\label{sect.3}
\subsection{Canonical model}
\label{Canonical model}
In this section we consider the cluster S0 at 12Gyr. This cluster is characterized by a central IMBH of $1.0\times 10^4\ M_\odot$ which represents $\sim 3$\% of the total mass (and it thus serves as a clear case to test detection in our study under the favourable conditions of a massive central IMBH). 

\begin{figure*}
\resizebox{\hsize}{!}{\includegraphics[clip=true]{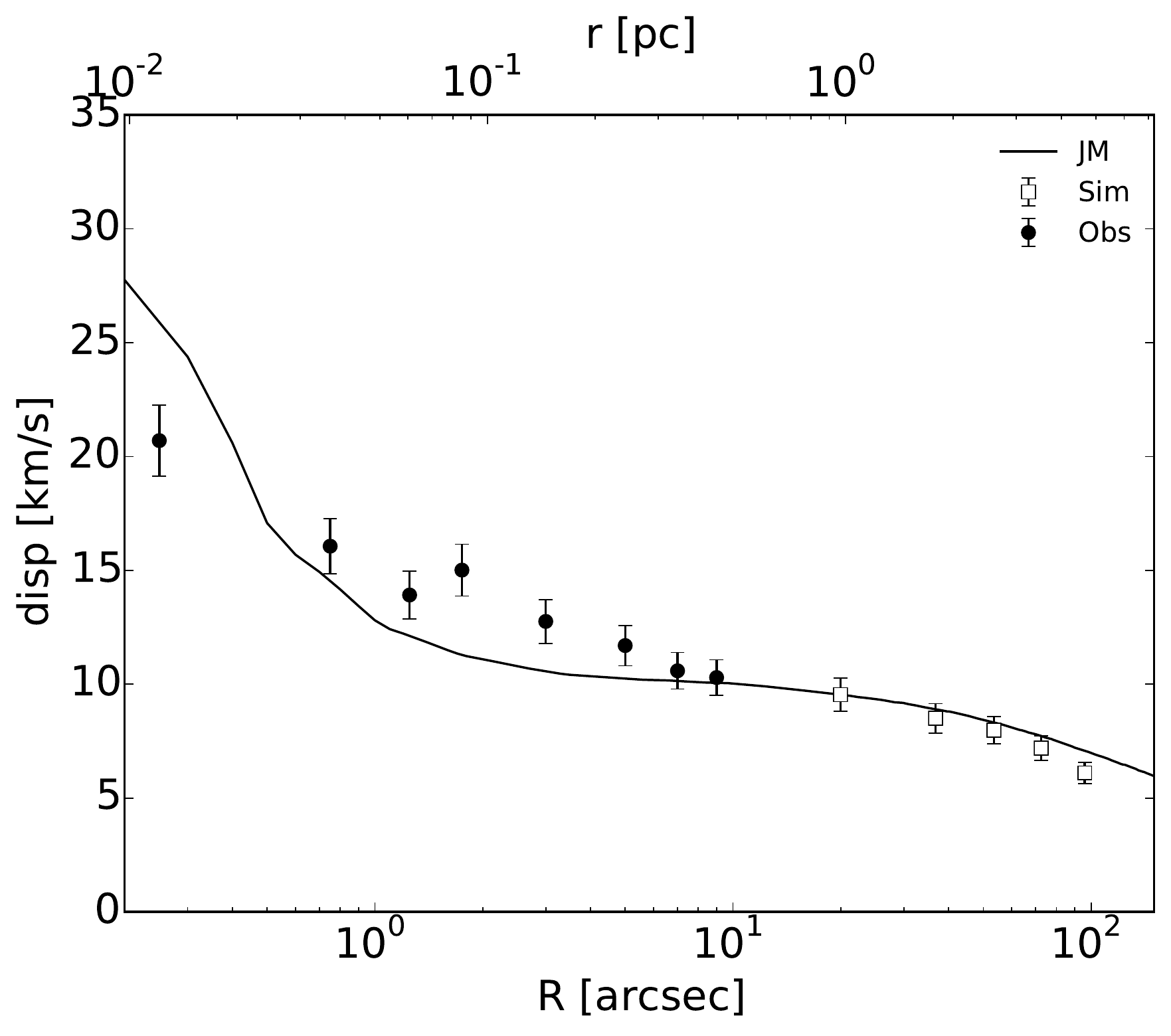}
\includegraphics[clip=true]{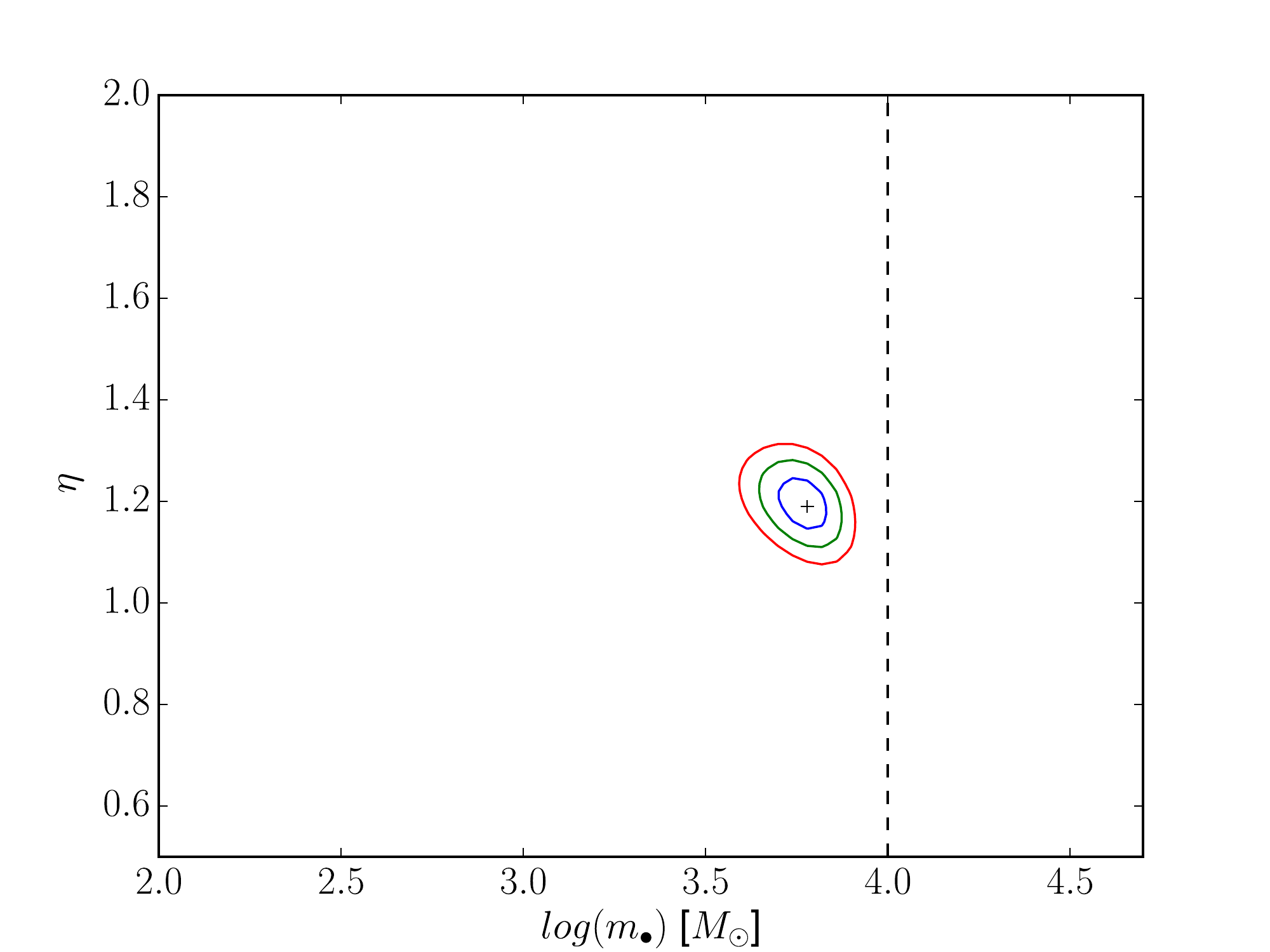}
}
\caption{{\it Left panel:} Best fit Jeans model for the cluster in simulation S0 at 12 Gyr. The observed inner profile (solid circles) is obtained at a distance of $10$ kpc pointing the FOV to the right centre of the cluster. The outer profile (open squares) is obtained directly from the simulation by considering only the velocities of the red giants. 
{\it Right panel:} 2D map of the chi-square function in the two parameters $\eta$ and $m_\bullet$. The three regions correspond to confidence levels of 68.3\%, 95.4\% and 99\%. The dotted line indicates the true mass of the IMBH. The Jeans model is not able to recover the mass of the IMBH ($10^4\ M_\odot$) in the 3-sigma confidence.
}
\label{bf}
\end{figure*}

At a distance of 10 kpc, a total of 38400 stars fall in the field of view (see the luminosity map in Fig.~\ref{LM+VDP}). The observed velocity dispersion profile is constructed by masking the IFU data following the procedure described in Sec.~\ref{subsect:vel disp prof}. Then, the chi-square function in Eq.~\eqref{chi2} is minimised in the two dimensional parameter space, giving a best fit value for the IMBH mass of  $6.0\substack{+0.6 \\ -1.0}\times10^3\ M_\odot$ and a value for $\eta$ of $1.19\substack{+0.04 \\ -0.04}$ (the errors are estimated with 68.3\% confidence). In the left panel of Fig.~\ref{bf} we plot the best fit Jeans model in comparison with the observed velocity dispersion profile. We also identify the regions corresponding to confidence levels of 68.3\%, 95.4\% and 99\% (see Fig.~\ref{bf}, right panel) finding that the true IMBH mass is higher by a factor $\approx1.25$ with $3\sigma$ confidence.

\subsection{Dependence on the IMBH mass}

\begin{figure*}
\resizebox{\hsize}{!}{\includegraphics[clip=true]{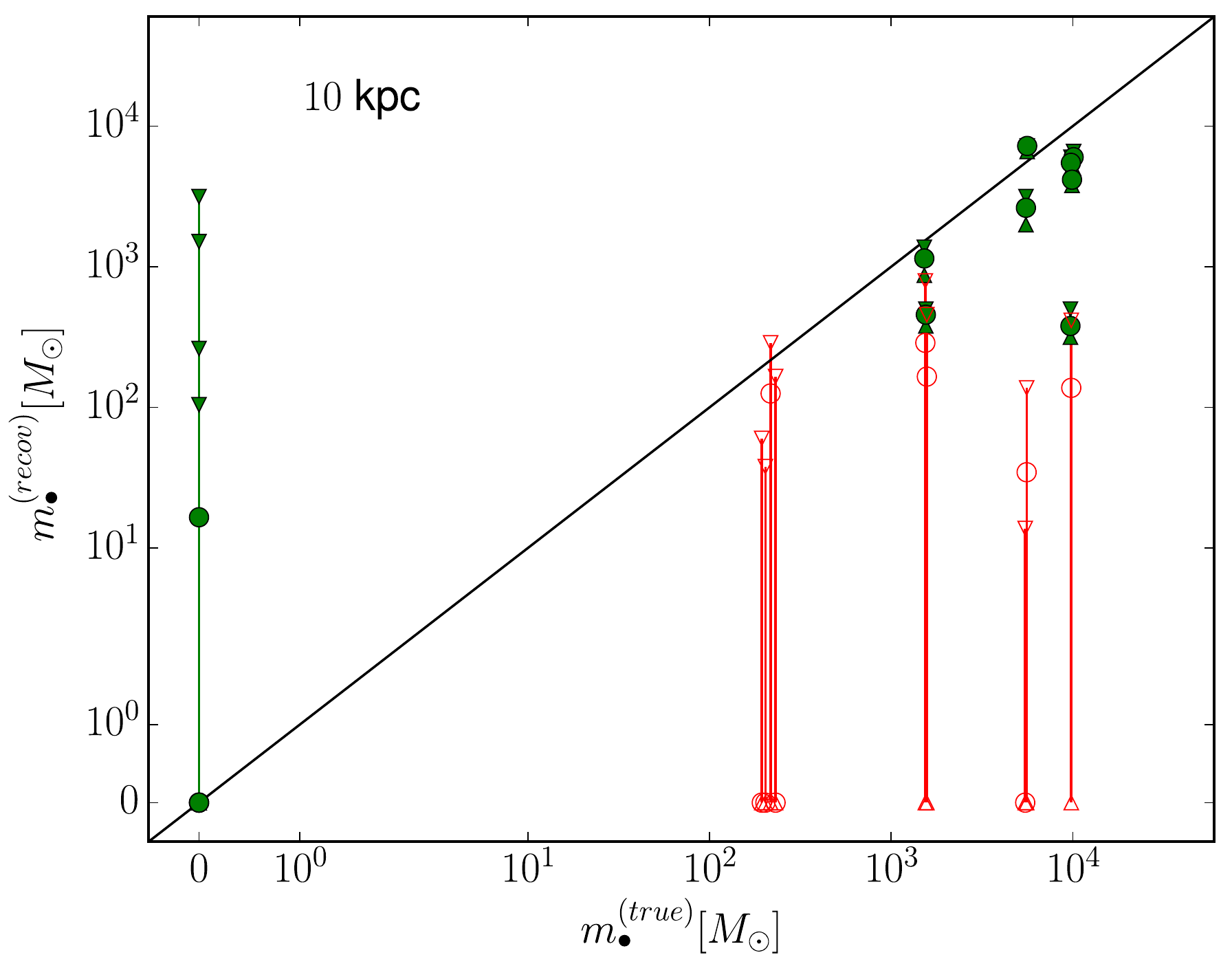}
\includegraphics[clip=true]{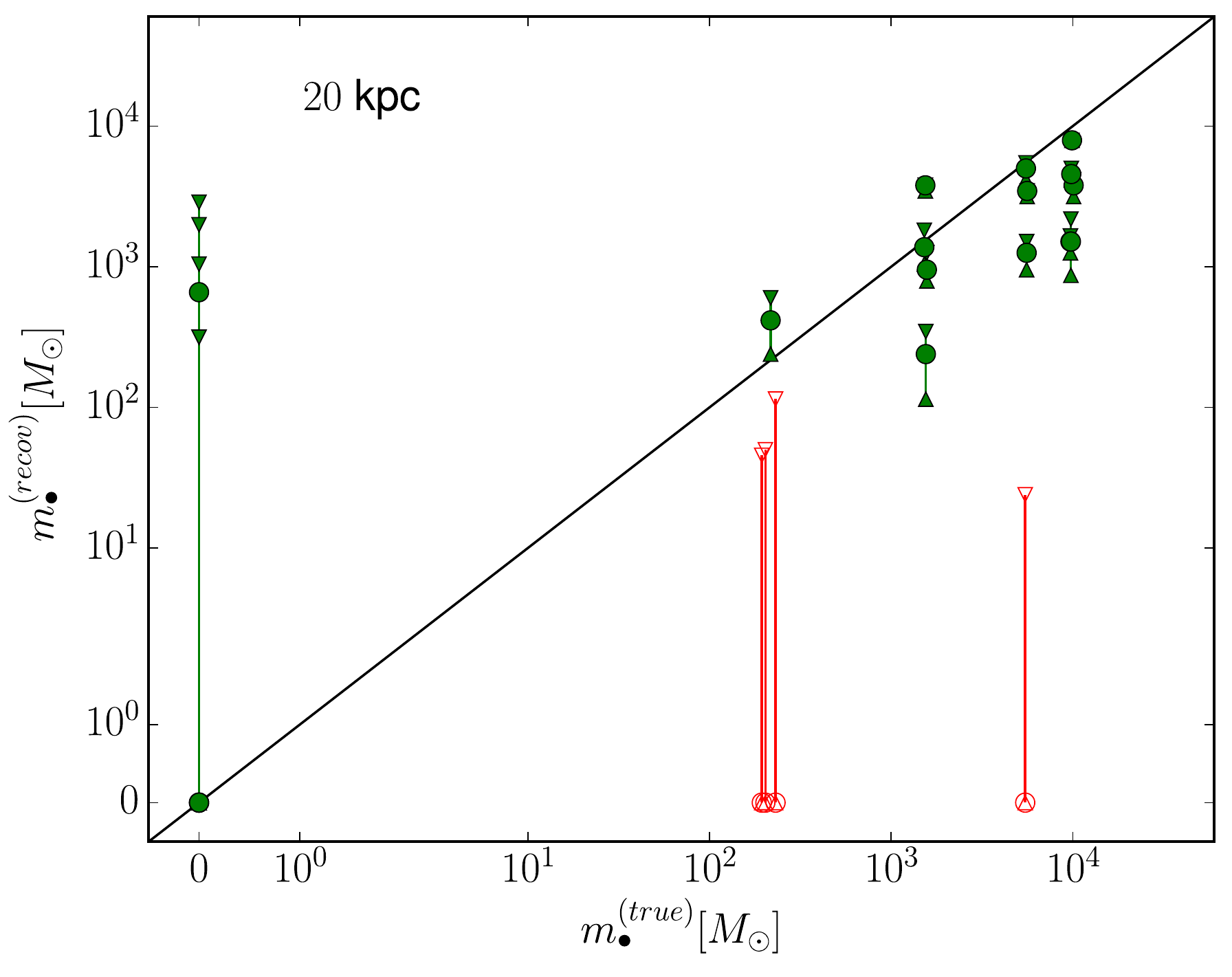}
}
\caption{Comparison between the true mass of the IMBH from the simulations and the mass recovered for the best fit of the mock observation. All the clusters are considered at a fixed distance of 10 kpc ({\it left panel}) and 20 kpc ({\it right panel}) by identifying the right centre for the FOV. The black line represents the relation $m_\bullet^{(true)}=m_\bullet^{(recov)}$. Every circle is the best fit mass $m_\bullet$ of the IMBH, while the error bars correspond to the confidence interval of $68.3\%$. As solid green circles we show the successful cases in which the observations are able to determine a lower limit greater than zero for the IMBH mass, while as open red circles we represent the cases in which the best fit mass is consistent with zero. For the simulation without the IMBH we plot in green those cases in which the best fit mass is consistent with zero within 1-sigma confidence.
}
\label{mtrueVSmrec}
\end{figure*}

As described in Sec.~\ref{simulations}, our set of simulations includes systems with a range of IMBH masses but otherwise similar properties. By applying the same analysis presented in Sec.~\ref{Canonical model} to the different models listed in Table~\ref{tab:SIM}, we aim to study the fidelity of the IMBH mass measurement as a function of mass itself. Qualitatively, we expect that, at fixed observational setup, the chances of recovering the IMBH mass correctly increase with mass (since the sphere of influence and the IMBH contribution to the central velocity dispersion profile are both larger). 

In Fig.~\ref{mtrueVSmrec} we show the IMBH mass recovered from the best fitting Jeans models as a function of the intrinsic IMBH mass. We present the results for each of the 4 snapshots (at 11.7, 11.8, 11.9 and 12 Gyr) of the simulations S1-S5 considered at a distance of 10 and 20 kpc (for the simulation S0 we considered only the snapshot at 12 Gyr). The results are color-coded according to the ability of the models to infer the presence of an IMBH with 1-sigma confidence. The red open circles represent the cases in which a solution without an IMBH is allowed at 1-sigma (or, for the simulation S5, in which a solution with  $m_\bullet>0$ is found at 1-sigma), and are flagged as failures. The green circles represent the cases identified as proper detections. This means that the best fit model is able to find a lower limit greater than 0 to the IMBH mass (or, in the case of the simulation S5 without the IMBH, a lower limit equal to 0). Note that these observations include cases in which the IMBH mass is greater than 0, but not consistent with the real mass of the simulation. These subcases are the majority in our sample as the mass inferred from the mock observations is found to generally underestimate the true mass. This might possibly happen because of systematic introduced by binary stars. The increment in the velocity dispersion due to the presence of binaries may exhibit radial variations which are not captured by our simple radial-independent correction factor $\eta$. Thus, our analysis may tend to overestimate the contribution of stars (and, subsequently, underestimate the contribution of the IMBH) to the central velocity dispersion. By taking into consideration this effect in a more refined modeling we would expect the green circles below the reference line to uniformly shift upwards and, eventually, intercept the true mass. However a more sophisticated treatment of the impact of binaries would rely on knowledge that is generally not available, nor used, in Jean-model analysis of actual observations, rather than mock data. Therefore, it would not be appropriate to implement such modeling to our mock IFU dataset. 

The expected trend with the IMBH mass is partially recovered (see Fig.~\ref{mtrueVSmrec}). Indeed, from an IMBH of few hundred $M_\odot$ to a high mass IMBH of $10^4\ M_\odot$ the successful probability increases from the 0\% to the 80\% of the cases considered at 10kpc. Also, we notice that, even for the high range of IMBH mass, some observations fail. This confirms how the stochasticity, which affects integrated-light measurements, arises even for a single cluster observed at different dynamical times. Finally, for the simulation S5 without the IMBH, all the cases at both distances are consistent with a cluster with no IMBH, even though the model is in general unable to exclude at 68\% confidence a IMBH with mass of $10^3 M_\odot$.

Besides the dependence on the IMBH mass, we are interested in studying how two other parameters can affect the probability of recovering the IMBH mass. In particular, we want to quantify the importance of identifying the right centre for the observation and to explore the dependence on the distance to the cluster. 

\subsection{The identification of the centre}

The identification of the centre in which to carry out the analysis is fundamental within the typical observational assumption of modeling in spherical geometry. In practical cases, there are two main sources of uncertainty associated with the centre. First, it is often challenging to determine the position of the cluster centre to high accuracy, as illustrated, for example, by the extensive debate on where the centre of $\omega$ Cen is (see, e.g, \citealt{noyola:10,vandermarel:10}). In addition, even when the centre of light (or the kinematic centre) of the system is identified with high-accuracy, there is no guarantee that the BH location coincides with it, especially in the case of a light IMBH (see e.g., \citealt{giersz:15}, \citealt{haster:16}).

\begin{figure*}
\resizebox{\hsize}{!}{\includegraphics[clip=true]{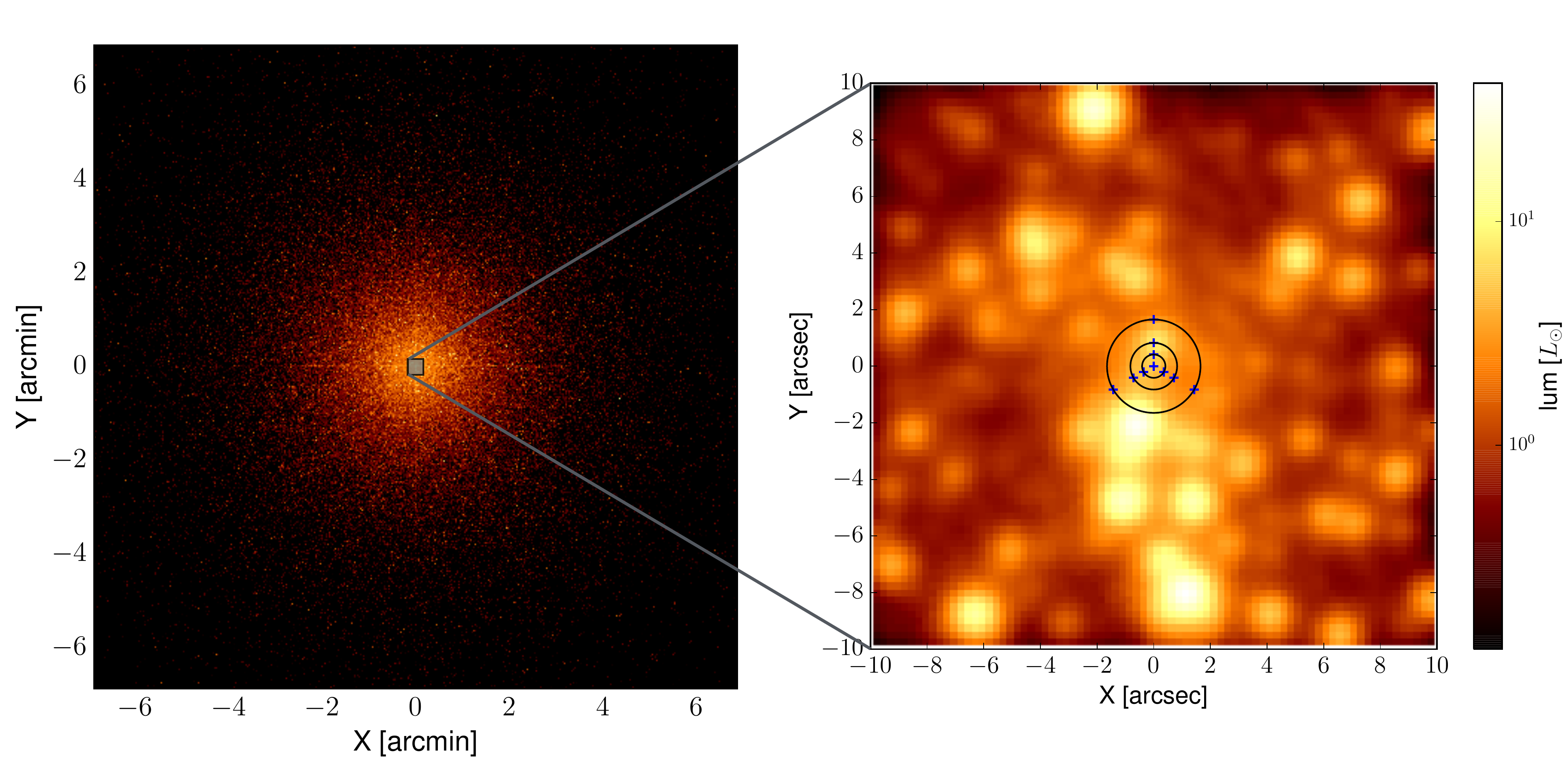}
}
\caption{{\it Left panel}: Luminosity map of the cluster S1 at a distance of 10 kpc. {\it Right panel}: detail of the $20\arcsec\times20\arcsec$ FOV. The blue crosses represent the 9 different centres of the analysis we investigate, distributed at three different radial positions with a 5\%, 10\% and 20\% off-sets with respect to the projected core radius from the intrinsic density centre of the simulation.
}
\label{Centres}
\end{figure*}

For two snapshots (at 11.7 and 12 Gyr) of the simulations S1 (high mass IMBH) and S3 (low mass IMBH), we analyse FOVs placed at different radial offsets of the centre (the distance to the clusters is fixed at 10 kpc). We consider a total of 9 different centres, three for each radial positions $0.4125\arcsec$, $0.825\arcsec$ and $1.65\arcsec$ corresponding for both clusters to radial offsets of 5\%, 10\% and 20\% of the core radius (see Fig.~\ref{Centres}). In terms of the sphere of influence of the central IMBH (whose radius $r_\bullet$ is defined as the radius at which the cumulative mass equals the IMBH mass), the offsets are in the range $0.03-0.14r_\bullet$ for the simulation S1 and in the range $0.08-0.3r_\bullet$ for the simulation S3.

\begin{figure*}
\resizebox{\hsize}{!}{
\includegraphics[clip=true]{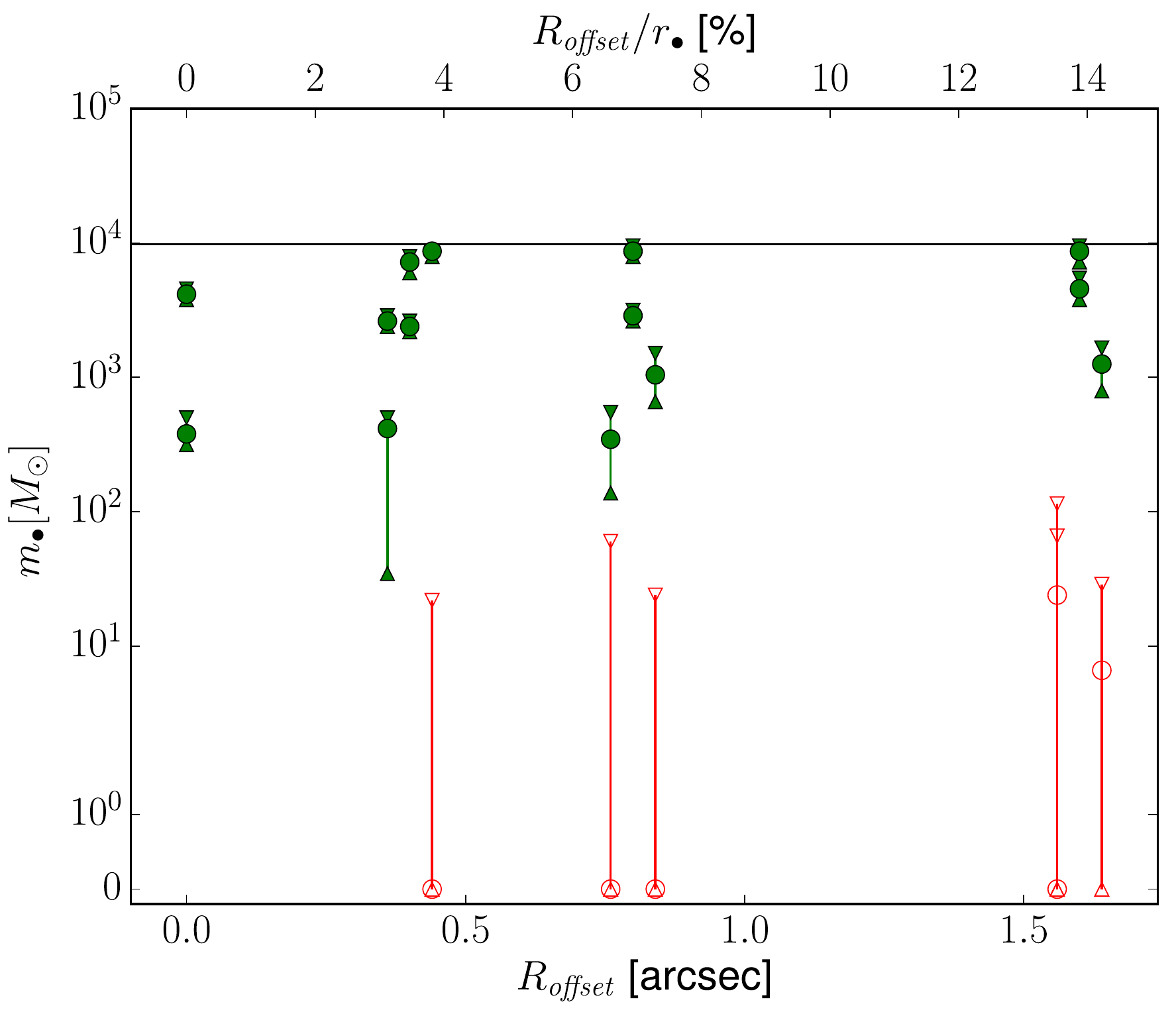}
\includegraphics[clip=true]{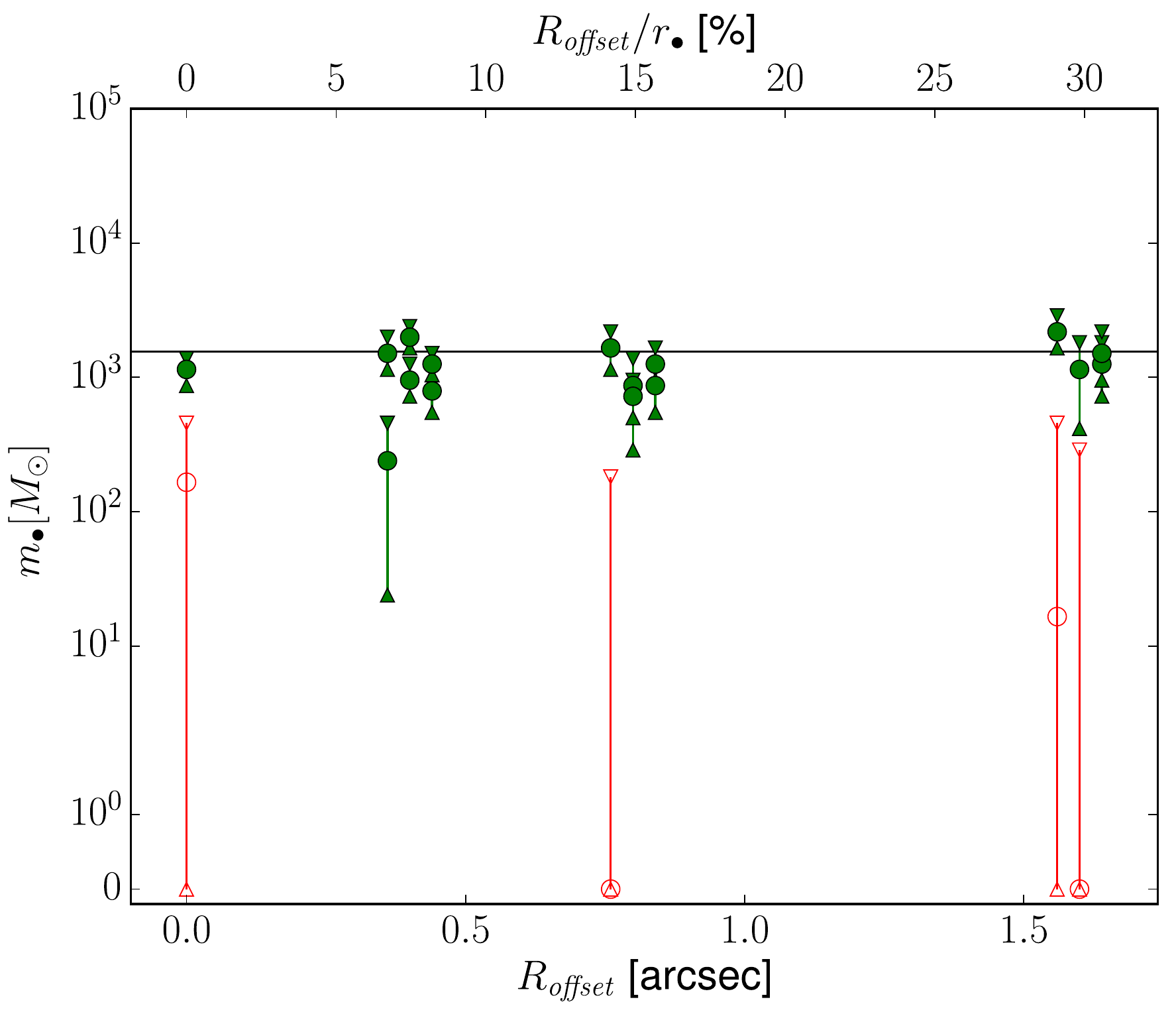}
}
\caption{
 Recovered IMBH mass for the 11.7 and 12 Gyr snapshots of the cluster S1 ({\it left panel}, high BH mass)  and the cluster S3 ({\it right panel}, low BH mass)  considered at a distance of 10 kpc (for a detailed descriptions of the symbols used see Fig.~\ref{mtrueVSmrec}). For every (positive) radial off-set there are 6 estimates of the mass, two (from different snapshots) for each of the three points equidistant from the centre (see Fig.~\ref{Centres}), which are shown with a slight shift along the x-axis in the figure for improving clarity. The probability of detecting an IMBH slightly decreases by increasing the off-sets.
}
\label{masscentre}
\end{figure*}

With the same notation used in the previous subsection we show the results of the analysis in Fig.~\ref{masscentre}. For both the clusters we analyse two different snapshots at 11.7 and 12 Gyr. In both cases the probability of detecting an IMBH slightly decreases by increasing the off-set. However, this trend is deeply influenced by the stochasticity as for a fixed radial off-set the success of the observation changes according to the angular position of the centre (e.g., see the high IMBH mass case with a fixed off-set of $1.6\arcsec$).

We wish to emphasise the fact that for our observations we are using all the information available from the simulation to produce the Jeans models used in the fitting procedure. The same modeling procedure is unavailable for real observations and, thus, we expect a wrong identification of the centre to reduce the successful probability found in our work (for a comparison with real observations, see the discrepancy of $\simeq8\%\ R_c$ for the centre of Omega Cen in \citealt{noyola:08} and \citealt{vandermarel:10}).  

\subsection{Changing the distance} 

Among the parameter that we change in our mock observations there is the distance to the cluster. The central IMBH is characterized by a sphere of influence that, in first approximation, depends only on its mass. Therefore, for one particular simulation and for a fixed resolution of the instrument, increasing the distance to the cluster has the same effect of reducing the sphere of influence of the black hole. As consequence, the central peak in the velocity dispersion is expected to reduce with increasing distance. As opposite effect, a more crowded FOV obtained by considering higher distances should limit discreteness effects such as the shot noise introduced by bright stars. 

\begin{figure*}
\resizebox{\hsize}{!}{
\includegraphics[clip=true]{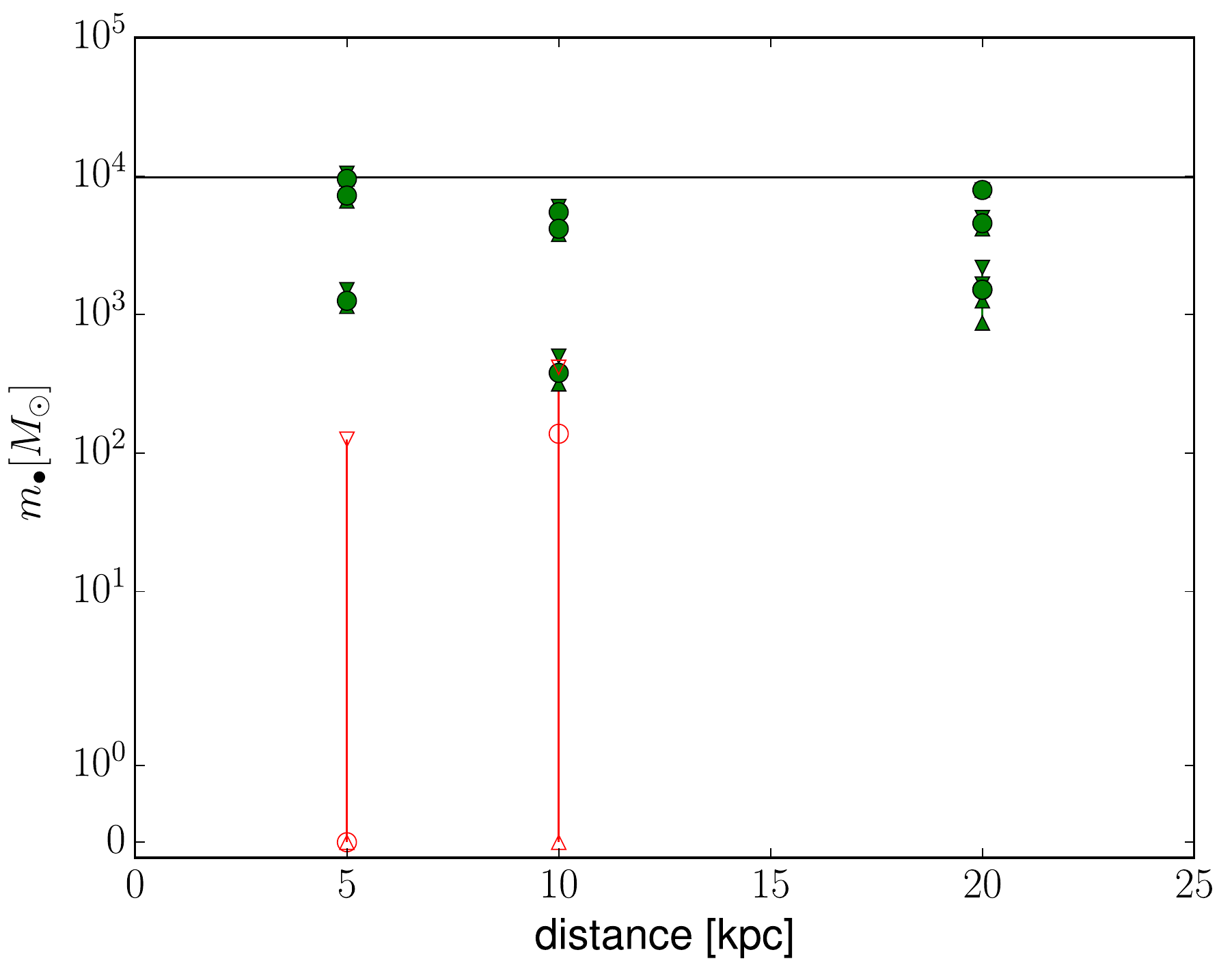}
\includegraphics[clip=true]{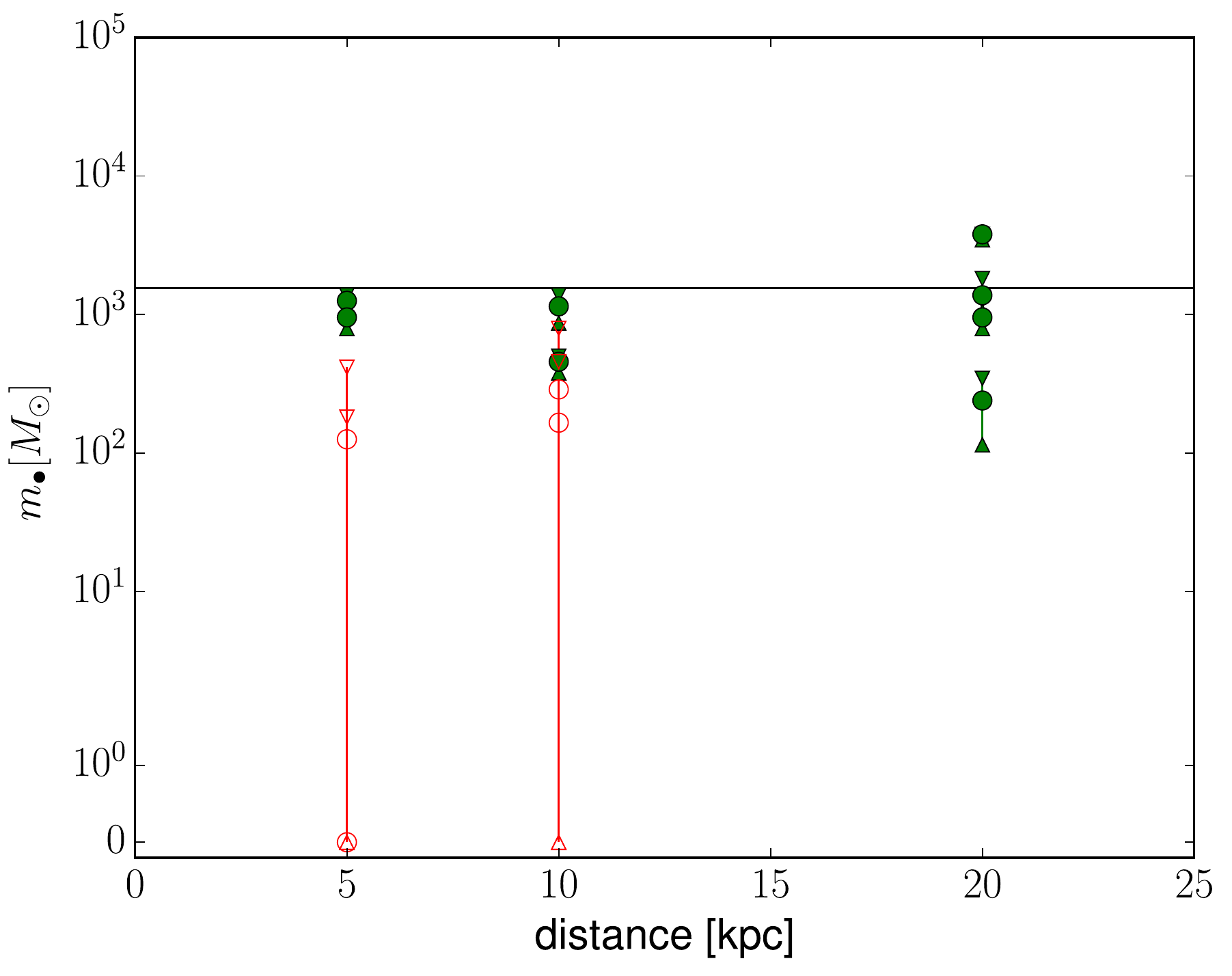}
}
\caption{Recovered IMBH mass as function of the distance to the sun for the cluster S1 ({\it left panel}, high BH mass) and S3 ({\it right panel}, low BH mass) at 11.7, 11.8, 11.9 and 12Gyr. The symbol used are the same of Fig.~\ref{mtrueVSmrec}.}
\label{massdistance}
\end{figure*}

We consider 3 different selected distances: 5, 10 and 20 kpc. In Fig.~\ref{massdistance}, we plot the recovered mass as a function of the distance to the cluster for each of the 4 snapshots available for the simulations S1 and S3. The probability of recovering an IMBH is marginally influenced by the distance to the cluster with the higher number of successes found at 20 kpc, where the influence of shot noise on the velocity dispersion profile is reduced.

\section{Conclusions}
\label{sect.4}
We simulated different integrated-light IFU observations for a sample of MOCCA simulations characterized by a series of realistic ingredients (high number of stars, stellar evolution, primordial binaries) and by different IMBH mass. Our goal was to test under which conditions the IMBH is recovered from the fit of a family of Jeans models to the mock observed velocity dispersion profile. 
We started by considering the different simulated clusters in canonical observational conditions, that is at a typical distance of 10 kpc and by identifying the right centre of the field of view. Even though we adopted an optimal masking procedure to limit the effect of the most bright stars, we find that our results are significantly influenced by the intrinsic stochasticity of the IFU measurements. Indeed, for every class of IMBH considered we found at least one snapshot in four for which the observed IMBH mass is consistent with the case without black hole. Decreasing the IMBH mass leads to a larger probability of failing to infer the BH presence from Jeans modeling, with probability of obtaining a null result going up to $\sim$75\% for IMBH mass 0.1\% of the total cluster mass. In addition, even when the IMBH presence is successfully recovered from the modeling, the inferred mass is systematically under-estimated, possibly because of the hard-to-quantify impact of binary stars (see Fig.~\ref{mtrueVSmrec}). Finally, in the large majority of snapshots without an IMBH, the Jeans modeling was not able to set reasonably low constraints to the inferred IMBH mass, that is even if the best fit mass was consistent with 0, it was impossible to exclude the presence of a massive IMBH at 68\% confidence.

In the second part of the paper we focused on changing crucial parameters of the observational setup. In particular, we explored the effects that the distance to the cluster and the centre of the field of view have to the inferred IMBH mass. The dependences on these two observational features have been analysed for two different regimes: high IMBH mass and low IMBH mass. In both cases we found similar trends. 

For a misidentification of the cluster centre not greater than the 5-20\% of the core radius we find that the presence of a central IMBH is successfully recovered in most observations. For both simulations, the highest number of failed observations corresponds to an off-set of $20\%\ R_c$ (corresponding to $\sim1.6\arcsec$ at $10$ kpc), suggesting a slight dependence of the successful probability with the centre off-set. We expect that this trend dramatically increases in real observations, for which the Jeans modeling is based on some assumptions on the light distribution.

Finally, the recovered IMBH mass is not particularly influenced by changing the distance between 5 and 20 kpc, even if the number of successes is higher at 20 kpc. According to the Harris catalogue (\citealt{harris:96}, 2010 edition) the 71\% of all the Galactic GCs are found in this range of distances.

Overall, our findings demonstrate that ground-based IFU observations of the cores of GCs can be very helpful tools to investigate whether IMBHs are present in galactic GCs, especially because it does not appear that increased distance induces a higher failure rate in the recovery of input IMBHs. However, a large sample of objects would be required in order to draw meaningful conclusions on the average IMBH occupation fraction in GCs. In fact, the failure rate of any single observation is high (25-100\% depending on BH mass) due to stochastic superposition of bright stars along the line of sight to the IMBH, and this bias needs to be corrected for. In conclusion, this work shows how any future IFU observation needs to be supported by other techniques with the purpose of providing complementary approaches. Even with their own observational limitations, either proper-motion based kinematics, such as that available from Hubble Space Telescope imaging at multiple epochs \citep{anderson:10,bellini:14}, or discrete kinematics from resolved-star spectroscopy \citep{lanzoni:13,kamann:16} may be used to constrain (especially in the outer regions) any tentative detection from integrated-light IFU observations . 


\section*{Acknowledgements}

This work was partially supported by the A.A.H. Pierce Bequest at the University of Melbourne and by the ``Angelo della Riccia'' grant to fund a visit of RdV to MPIA. PB acknowledges partial support from a CITA National Fellowship. AA and MG were partially supported by the Polish National Science Centre (PNSC) through the grant DEC-2012/07/B/ST9/04412. AA would also like to acknowledge partial support by the PNSC through the grant UMO-2015/17/N/ST9/02573 and partial support from NCAC grant for young researchers. GvdV acknowledges partial support from Sonderforschungsbereich SFB 881 "The Milky Way System" (subproject A7 and A8) funded by the German Research Foundation.




\bibliographystyle{mnras}
\bibliography{biblio} 



\bsp	
\label{lastpage}
\end{document}